\documentclass[conference]{IEEEtran}
\IEEEoverridecommandlockouts
\usepackage{amsmath,amssymb,amsfonts}
\usepackage[utf8]{inputenc}
\usepackage{cite}
\usepackage{graphicx}
\usepackage{textcomp}
\usepackage{url}
\usepackage[caption=false]{subfig}
\usepackage[justification=centering]{caption}
\usepackage{epsfig}
\usepackage{float}
\usepackage{xcolor}
\usepackage{balance}
\usepackage{bbm}
\usepackage{textcomp}
\usepackage{enumitem}
\usepackage{mathtools}
\usepackage[ruled,vlined]{algorithm2e}

\ifCLASSINFOpdf
\else
\fi
\newcommand{\linebreakand}{%
  \end{@IEEEauthorhalign}
  \hfill\mbox{}\par
  \mbox{}\hfill\begin{@IEEEauthorhalign}
}

\begin{document}
\title{Power Minimizing MEC Offloading with QoS Constraints over RIS-Empowered Communications\thanks{The work of F. Costanzo, K. D. Katsanos, G. C. Alexandropoulos, and P. Di Lorenzo was supported by the EU H2020 RISE-6G project under grant number 101017011.}}

\author{Mattia Merluzzi$^1$, Francesca Costanzo$^2$, Konstantinos D. Katsanos$^4$, \\George C. Alexandropoulos$^4$, and Paolo Di Lorenzo$^{2,3}$\\
$^1$CEA-Leti, Université Grenoble Alpes, F-38000 Grenoble, France\\
$^2$DIET department, Sapienza University of Rome, Italy, $^3$CNIT, Parma, Italy
\\
$^4$Department of Informatics and Telecommunications, National and Kapodistrian University of Athens, Greece
\\
emails: mattia.merluzzi@cea.fr,\{francesca.costanzo,paolo.dilorenzo\}@uniroma1.it,\{kkatsan,alexandg\}@di.uoa.gr
\vspace{-.3 cm}}
\maketitle
\begin{abstract}
This work lies at the intersection of two cutting edge technologies envisioned to proliferate in future 6G wireless systems: Multi-access Edge Computing (MEC) and Reconfigurable Intelligent Surfaces (RISs). While the former will bring a powerful information technology environment at the wireless edge, the latter will enhance communication performance, thanks to the possibility of adapting wireless propagation as per end users' convenience, according to specific service requirements. We propose a joint optimization of radio, computing, and wireless environment reconfiguration through an RIS, with the goal of enabling low power computation offloading services with reliability guarantees. Going beyond previous works on this topic, multi-carrier frequency selective RIS elements' responses and wireless channels are considered. This opens new challenges in RIS optimization, accounting for frequency dependent RIS response profiles, which strongly affect RIS-aided wireless links and, as a consequence, MEC service performance. We formulate an optimization problem accounting for short and long-term constraints involving device transmit power allocation across multiple subcarriers and local computing resources, as well as RIS reconfiguration parameters according to a recently developed Lorentzian model. Besides a theoretical optimization framework, numerical results show the effectiveness of the proposed method in enabling low power reliable computation offloading over RIS-aided frequency selective channels. 
\end{abstract}
\vspace{0.05 cm}
\begin{IEEEkeywords}
Multi-access Edge Computing, Reconfigurable Intelligent Surfaces, Energy-efficient wireless networks, 6G
\end{IEEEkeywords}

%
\IEEEpeerreviewmaketitle
\section{Introduction}
Today, as part of the race to 6G, communications and computing are converging towards a unified complex system in which data are continuously exchanged among heterogeneous intelligent agents that operate computational demanding operations such as training, data distillation, and inference \cite{Letaief2022}, over severe wireless channels. This explosion of data requires an unprecedented effort in conceiving wireless systems with a unified perspective in which computing is a natural component of the network, to be designed, optimized, and operated jointly with communication. To this end, in the last few years, Multi-access Edge Computing (MEC) \cite{kekki2018etsi} has been identified as a key technological enabler of this vision, thanks to its fundamental characteristic of bringing such resources close to the end service consumers, thus enabling low-latency, energy-efficient, and dependable connect-compute operations. To enhance dependability aspects, Reconfigurable Intelligent Surfaces (RISs) \cite{huang2019reconfigurable,huang2020holographic} have been identified as a key component of 6G and beyond \cite{RISE6G_COMMAG}, thanks to their ability of opportunistically and dynamically controlling signal reflections, to enhance reliability and diversity, thus avoiding unexpected detrimental service disruptions and boost the performance in intended areas across space, time, and frequency \cite{rise6g}. In this paper, we focus on the specific case of computation offloading services in MEC and RIS-enabled networks, exploring the promising marriage of these technologies towards the 6G vision.\\
\textbf{Related works.} The research efforts on RISs and MEC-aided systems have already started, with contributions on edge caching \cite{chen2021exploiting} and computation offloading \cite{bai2020latency,chu2020intelligent,huang2021reconfigurable,hu2021reconfigurable}. Specifically, the authors of \cite{bai2020latency} propose to minimize the latency in a multi-user scenario, optimizing the computation offloading volume, the edge computing resource allocation, the multi-user detection matrix, and the RIS configuration, under the constraint of a total edge computing capability. In \cite{chu2020intelligent}, the authors maximize the number of processed bits for computation offloading, optimally designing the edge server, or Mobile Edge Host (MEH) CPU frequency, the offloading time, the transmit power, and the RIS configuration. Also, \cite{huang2021reconfigurable} exploits RISs to maximize the performance of a machine learning task run at the MEH, acting jointly on radio parameters such as User Equipment (UE)'s power, Access Point (AP) beamforming, and RIS phase shifts. Furthermore, data-driven solutions for RIS-empowered multi-user mobile edge computing is proposed in \cite{hu2021reconfigurable}, with the goal of maximizing the total completed task-input bits of all UEs under energy budget constraints. Finally, in \cite{di2022dynamic}, a joint optimization of radio, computing, and RIS parameters was proposed considering frequency-flat channels and RIS responses. A similar contribution was presented in \cite{airod2022} for the case of multiple-input-multiple-output communications. None of these works considers a realistic frequency selective model of channels and RISs' element response \cite{Mancera2017_all,Katsanos_Lorentzian}.\\
\textbf{Our contribution.} In this work, we propose a resource allocation algorithm to dynamically optimize radio, computing, and frequency dispersive aware RIS configuration, over multi-carrier frequency selective channels. Following \cite{Mancera2017_all}, we model the elements' response with a Lorentzian form, and we optimize it accordingly. Our goal is to seek for the minimum power consumption of a user performing local computation before uploading intermediate results to a MEH, through the wireless connection with an AP, under average and probabilistic delay constraints, with no prior knowledge of the statistics of time varying computational demands, wireless channels, MEH CPU availability, and offloaded tasks' sizes. After formulating a long-term problem, we decouple it into a sequence of deterministic problems to be solved in a per-slot fashion, only based on instantaneous observations. The resulting sub-problems are then solved with low-complexity heuristics and closed forms when possible, with QoS guarantees. Numerical results show the effectiveness of our method and the convenience of using a frequency selective aware dynamic RIS configuration when optimizing connect-compute resources. To the best of our knowledge, this is the first work merging RIS and MEC technologies over frequency selective channels and RIS element responses.\\
\textit{Notation:} Bold lower and upper case letters denote vectors and matrices, respectively; 
the operator $|\cdot|$ denotes the absolute value of a complex number, while $card(\cdot)$ denotes the cardinality of a set. Finally, given a random variable $X$, its long-term average is denoted as $\overline{X}$
\begin{equation}\label{long_term_avg}
    \overline{X}=\lim\nolimits_{T\to\infty}\frac{1}{T}\sum\nolimits_{t=1}^T\mathbb{E}\{X_t\}
\end{equation}

\section{System Model and Assumptions}
Let us consider a scenario in which time is organized in slots $t=1,2,3,\ldots$, with an end device (e.g. a UE or a sensor) periodically generating new computation tasks to be executed under targeted service performance. Such tasks entail a local computation demand, and a remote computation portion to be offloaded to the MEH, as well as a number of bits needed to enable such remote processing (i.e. the offloading task size). Then, the end-to-end (E2E) delay comprises local computing, transmission, and remote computation. 
\subsection{RIS's frequency response and channel model}\label{subsec:RIS_model}
An RIS is generally composed of nearly passive elements, whose phases can be opportunistically tuned. However, while most of the existing models exploited in the literature consider the RIS response as constant across frequency, we model the system according to \cite{Katsanos_Lorentzian}, where the resonant elements (described as inductor-capacitor resonators in circuit theory) of an RIS exhibit a frequency dispersive property. By assuming electrically small metamaterial elements to implement the RIS, we can model each $n$-th element as a polarizable dipole, whose frequency response takes the following Lorentzian form \cite{Mancera2017_all}:
\begin{equation}
	\label{eqn:Lorentz}
	\phi_n(x) \triangleq  \frac{S_n x^2}{f_n^2 -x^2+j\kappa_n x},\quad n=1,2,\ldots,N,
\end{equation}
where $S_n$, $f_n$, and $\kappa_n$ are the element-dependent oscillator strength, resonance frequency, and damping factor, which can be externally controlled. According to \cite{WangHanqing2021}, the damping factor can be written as $\kappa_n\triangleq\frac{f_n}{2\chi_n}$, where $\chi_n$ is a tunable quality factor. The quality factor 
determines the bandwidth influenced by an RIS element. The smaller is $\chi_n$, the larger is the influenced bandwidth, and vice versa \cite{Katsanos_Lorentzian}. 
Let us consider a multi-carrier system with subcarrier spacing $W$, and central frequency $f_b$ on frequency bin $b\in\mathcal{B}$, with $\mathcal{B}$ the set of bins. Then, the Lorentzian in \eqref{eqn:Lorentz} takes the following discrete form on subcarrier $f_b$ at time slot $t$
\cite{Smith17,WangHanqing2021}:
\begin{equation}
	\label{eqn:Lorentz1}
	\phi_{n,t}(f_b) =  \frac{S_{n,t} f_b^2}{f_{n,t}^2 -f_b^2+j \frac{f_{n,t}}{2 \chi_{n,t}} f_b},
\end{equation}
with $n=1,\ldots N$ the RIS element index, and $t$ the slot index. The time dependency shows that RIS reflection parameters can be optimized dynamically.
We denote by $h^{\rm LoS}_t(f_b)\in\mathbb{C}^1$ the frequency response of the direct channel between a single-antenna device and a single-antenna AP (receiver), while we denote by $\mathbf{g}_t(f_b)\in\mathbb{C}^{1\times N}$ and $\mathbf{h}_t(f_b)\in\mathbb{C}^{N\times 1}$ the vectors of frequency responses between the RIS and the AP, and between the device and the RIS, respectively. Also let us denote by  $\mathbf{\Phi}_t(f_b)\in\mathbb{C}^{N\times N}$, a diagonal matrix whose $n$-th diagonal elements is given by \eqref{eqn:Lorentz1}, and which varies with $b$ due to the frequency dispersive property of the RIS. Then, the overall channel to noise ratio on frequency bin $b$ can be written as follows:
\begin{equation}\label{CNR}
    \alpha_t(f_b)\triangleq\frac{|h^{\rm LoS}_t(f_b)+\mathbf{g}_t(f_b)\mathbf{\Phi}_t(f_b)\mathbf{h}_t(f_b)|^2}{N_0 W},
\end{equation}

\subsection{Local Computation Delay and Power Consumption}
We denote by $w^l_t$ the local computing demand (in CPU cycles), which varies over time, according to a possibly unknown distribution. Also, the device selects, at each time slot, its CPU clock frequency $f^l_t$ (in CPU cycles per second). Then, the time needed to process the local computation demand at slot $t$ is
\begin{equation}\label{local_comp_delay}
    D^l_t\triangleq w^l_t/f^l_t.
\end{equation}
Also, given $f^l_t$, we adopt the widely used cubic model for the CPU power consumption \cite{Yuan06}:
\begin{equation}\label{local_energy}
    p^l_t=\gamma(f^l_t)^3,
\end{equation}
where $\gamma$ is the effective switched capacitance of the processor.
\subsection{Uplink Transmit Power and Delay}
We denote by $p_{b,t}$ the portion of the user transmit power that is dynamically allocated on each bin $b$ at each time slot $t$. Denoting by $p_{\max}$ the maximum user transmit power over all $card(\mathcal{B})$ subcarriers, the following constraint must hold:
\begin{equation}\label{uplink_energy}
    p^u_t\triangleq\sum\nolimits_{b\in\mathcal{B}}p_{b,t}\leq p_{\max},
\end{equation}
where $p^u_t$ denotes the total power spent by the user to upload data, at slot $t$. Given \eqref{eqn:Lorentz1} and \eqref{CNR}, and assuming an OFDM system, we can write the total experienced uplink data rate at time slot $t$ through the Shannon formula:
\begin{equation}\label{data_rate}
    R_t\triangleq W\sum\nolimits_{b\in\mathcal{B}}\log_2\left(1+\alpha_t(f_b) p_{b,t}\right).
\end{equation}
Finally, denoting by $A_t$ the number of bits to be transmitted at time $t$ to enable the remote execution part, the uplink transmission delay simply reads as:
\begin{equation}\label{uplink_delay}
    D^u_t\triangleq A_t/R_t.
\end{equation}



\subsection{Remote Computation Delay and Energy Consumption}
At the MEC infrastructure side, we assume that, at each slot $t$, the MEH assigns a portion $\sigma_t\in(0,1]$ (with unknown statistics) of its total CPU cycle frequency $f_{\max}$ to the user, due, e.g., to higher priority traffic, which we consider as an exogenous variable. Then, the time needed to execute the remote computation demand $w^r_t$, at time slot $t$, is 
\begin{equation}\label{comp_delay}
  D^r_t=w^r_t/(\sigma_t f_{\max}),
\end{equation}
where the required amount $w^r_t$ of CPU cycles is a random variable with unknown statistics.

Finally, given \eqref{local_comp_delay}, \eqref{uplink_delay}, and \eqref{comp_delay}, the E2E delay reads as follows:
\begin{equation}\label{total_delay}
    D^{\textrm{tot}}_t=D^l_t+\left(D^u_t+D^r_t\right).
\end{equation}
Similarly, recalling \eqref{local_energy}, and \eqref{uplink_energy}, we can write the user power consumption as
\begin{equation}\label{energy_tot_user}
    p^{\textrm{tot}}_t=p^l_t+p^u_t.
\end{equation}

\subsection{Average Delay and Outage Probability}
In this work, we consider two types of service requirements: i) Average E2E delay; ii) Outage probability, i.e. the probability that the E2E delay exceeds a predefined threshold. The first requirement can be formalized as follows (cf. \eqref{long_term_avg}):
\begin{equation}\label{avg_const}
    \overline{D^{\textrm{tot}}}\leq D^{\rm avg}.
\end{equation}
However, only guaranteeing a bounded average E2E delay is not usually enough to meet service requirements, as long tails of the delay distribution can be highly detrimental for most of edge applications. To this end, we assume that each generated task is also required to be completed before a predefined delay threshold $D^{\max}$ with a certain probability, requested a priori by the user. 
More specifically, we assume an \textit{outage event} to occur if  $D^{\textrm{tot}}>D^{\max}$, 
while its probability, which we refer to as \textit{outage probability} in the sequel, is (cf. \eqref{long_term_avg})
\begin{equation}\label{prob_const}
    P^o\triangleq \textrm{Pr}\left\{D^{\textrm{tot}}_t>D^{\max}\right\}=\overline{u\left\{O_t\right\}} \leq \epsilon
\end{equation}
where $u\{\cdot\}$ denotes the unitary step function, $O_t=D^{\textrm{tot}}_t-D^{\max}$, and $\epsilon$ is a predefined threshold for outage tolerance. 
\section{Problem formulation and Solution}
The aim of this work is to design a computation offloading policy that involves radio, computing, and RIS optimization, to minimize user's power consumption under average delay and outage probability constraints. To this end, recalling \eqref{long_term_avg} and \eqref{energy_tot_user}, we formulate the following long-term problem:
\begin{align}\label{problem_form}
      &\hspace{0 cm}\min_{\{\mathbf{\Theta}_t\}_t}\; \overline{p^{\textrm{tot}}} \\
   &\textrm{subject to} \quad \eqref{avg_const},\eqref{prob_const};\nonumber\\
   &(a)\; 0 \leq S_{n,t}\leq 1,\;\forall n,t\quad\;\;\,
   (b)\;f_{n,t}\in\Omega,\; \forall n,t;\nonumber\\
   &(c)\;\chi_{n,t}\in\mathcal{X},\hspace{0.4cm}\forall n,t; \qquad (d)\; \; \left|\phi_{n,t}(f_b)\right|\leq 1,\;\forall n,b,t;\nonumber\\
   &(e)\;p_{b,t}\geq 0,\;\forall b,t\qquad\qquad(f)\;p_{\min}\leq\sum_{b\in\mathcal{B}}p_{b,t}\leq p_{\max},\;\forall t\nonumber\\
   &(g)\;f^l_{\min} \leq f^l_t\leq f^l_{\max},\;\forall t\nonumber
\end{align}
with $\mathbf{\Theta}_t=[\{f_{n,t}\}_{n},\{S_{n,t}\}_{n},\{\chi_{n,t}\}_{n},\{p_{b,t}\}_{b},f^l_t].$
Besides the long-term constraints \eqref{avg_const} and \eqref{prob_const}, the instantaneous feasibility constraints in \eqref{problem_form} have the following meaning: $(a)$ the parameter $S_{n,t}$ is chosen between $0$ and $1$, for each element of the RIS; $(b)$ the resonance frequency $f_{n,t}$ is chosen from a discrete set $\Omega$ of candidate values; $(c)$ the quality factor $\chi_{n,t}$ is chosen from a discrete set $\mathcal{X}$ of candidate values; $(d)$ the amplitude response of each RIS's element cannot exceed $1$ across all subcarriers; $(e)$ the transmit power allocated on each bin is non negative; $(f)$ the total transmit power is chosen between a minimum value $p_{\min}$ and a maximum value $p_{\max}$; $(g)$ the local CPU frequency is chosen between a minimum value $f_{\min}^l$ and a maximum value $f_{\max}^l$. Problem \eqref{problem_form} is extremely complex, due to the absence of statistical knowledge of context parameters involving wireless channels, computational demands, task offloading sizes, and MEH's CPU availability. Furthermore, the discrete nature of several involved variables makes it hardly tractable over the considered long-term horizon.
\subsection{The Lyapunov approach}
Lyapunov stochastic optimization is a powerful tool to transform long-term problems into a sequence of lower complexity deterministic problems, with convergence and asymptotic optimality guarantees with respect to the original problem. This is possible thanks to the definition of suitable state variables that are exploited to guarantee bounded time-averages of the involved variables (i.e. \eqref{avg_const}, \eqref{prob_const} in this case) \cite{Neely10}. 
More specifically, to deal with long-term constraints, we make use of mathematical models known as \textit{virtual queues}, able to track the state of the system in terms of constraint violations, to then take control actions, thus driving the system towards efficient and reliable operations. More specifically, to deal with \eqref{avg_const}, we define a virtual queue $Y_t$ that, between two consecutive time slots, evolves as follows:
\begin{equation}\label{Y_evolution}
    Y_{t+1}=\max\left(0,Y_t+D^{\textrm{tot}}_t-D^{\rm avg}\right).
\end{equation}
Similarly, for constraint \eqref{prob_const}, we define virtual queue $Z_t$, which evolves as follows
\begin{equation}\label{Z_evolution}
    Z_{t+1}=\max\left(0,Z_t+u\left\{O_t\right\}-\epsilon\right).
\end{equation}
The rationale behind the virtual queues is straightforward. Each virtual queue increases whenever the corresponding constraint is instantaneously violated, and decreases otherwise. Building on this, we can formally and theoretically guarantee constraints \eqref{avg_const}-\eqref{prob_const}, by guaranteeing virtual queues' mean rate stability\footnote{For a virtual queue $G_t$, it is defined as $\lim_{T\to\infty}\mathbb{E}\{G_T\}/T=0$.}. This transforms \eqref{problem_form} into a pure stability problem. To achieve mean rate stability of the virtual queues (i.e. to guarantee the long-term constraints) let us define the Lyapunov function \cite{Neely10}, $L(\mathbf{Q}_t)= \frac{1}{2}\left(Y^2_t+Z^2_t\right)$,
with $\mathbf{Q}_t=[Y_t,Z_t]$, which is a measure of the instantaneous congestion state of the system. Intuitively speaking, our goal is to push the system towards low congestion states (i.e. achieving stability) with the least power consumption (i.e. the objective function of \eqref{problem_form}). To this end, let us define the conditional Lyapunov drift (LD) $\Delta(\mathbf{Q}_t)=\mathbb{E}\left\{ L(\mathbf{Q}_{t+1})- L(\mathbf{Q}_t)|\mathbf{Q}_t\right\}$,
which is the conditional expected change of the Lyapunov function over successive slots, with the expectation typically taken with respect to context parameters (in this case wireless channels, computational demands, tasks' size, and MEH's CPU availability). A bounded LD leads to virtual queues' stability (i.e. long-term constraints guarantees) \cite{Neely10}. However, the LD does not assign any importance to the objective function of the original problem \eqref{problem_form}, i.e. the device power consumption. To tackle with both stability and power minimization, we define the \textit{drift-plus-penalty} (DPP) function as follows \cite{Neely10}:
\begin{equation}\label{DPP}
    \Delta_p(\mathbf{Q}_t) = \Delta(\mathbf{Q}_t)+V\mathbb{E}\left\{p^{\textrm{tot}}_t|\mathbf{Q}_t\right\}.
\end{equation}
The DPP is an augmented version of the conditional Lyapunov drift, which weights the drift and the objective function of the problem, thus shaping the trade-off between E2E delay guarantees (average and probabilistic) and device power consumption. Now, the same theoretical results apply to the DPP, meaning that by ensuring its bounded value at each slot, we guarantee the mean rate stability of the virtual queues, but also a power consumption that decreases as $V$ increases. To this end, following \cite{Neely10}, we now proceed by defining a suitable upper bound of the DPP, to be minimized afterwards with theoretical guarantees. In this case, the upper bound reads as 
\begin{align}
    \Delta_{p,t}\leq\zeta+\mathbb{E}\{&Y_t(D^{\textrm{tot}}_t-D^{\textrm{th}})+Z_t(u\left\{O_t\right\}-\epsilon)+ V p^{\textrm{tot}}_t|\mathbf{Q}_t\},\nonumber
\end{align}
with $\zeta$ a finite positive constant, omitted due to the lack of space, along its derivation and that of the upper bound, which follow similar arguments as in \cite{Neely10}.
Finally, hinging on the concept of opportunistically minimizing an expectation, we proceed by greedily minimizing the above upper bound at each time slot, thus formulating the following deterministic per-slot problem, which only requires instantaneous knowledge of the involved random variables:
\begin{align}\label{slot_prob}
      &\hspace{0 cm}\min_{\mathbf{\Theta}_t} \quad   V p^{\textrm{tot}}_t+Y_tD^{\textrm{tot}}_t + Z_t\mathbf{1}\left\{D_t^{\textrm{tot}}-D^{\max}\right\}  \\
   &\textrm{subject to} \; (a)\text{-}(g)\;\text{of} \;\eqref{problem_form}.\nonumber
\end{align}
Problem \eqref{slot_prob} is a mixed integer non linear, non differentiable program, due to the presence of the unitary step function, and the non convex nature of the data rate with respect to the RIS parameters. However, let us hinge on the upper bound $u\{D^{\textrm{tot}}_t-D^{\max}\}\leq D^{\textrm{tot}}_t+D^{\max}(\eta-1)$, with $\eta\geq 1/D_{\max}$, to approximate the unitary step function. This approximation dramatically simplifies \eqref{slot_prob}, however preventing to achieve its optimal solution at time $t$. Nevertheless, we hinge on the concept of $C$-additive approximation, which allows inexact solution of the per-slot problem, provided that their distance from the optimal one is bounded by a finite constant $C$ \cite{Neely10}. Now, given the above approximation, it can be easily shown that the problem can be split into a radio resource allocation sub-problem, involving power allocation and RIS configuration, and a computation resource allocation sub-problem to optimize local computing resources.
\subsection{Local computation sub-problem}
To allocate local computing resources, it is necessary to solve the following problem:
\begin{align}\label{slot_prob_local_comp}
      \hspace{0 cm}\min_{f^l_t} \; &V \gamma (f^l_t)^3+\left(Y_t+Z_t\right)w^l_t/f^l_t \\
   &\textrm{subject to}\quad f^l_{\min}\leq f^l_t\leq f^l_{\max}.\nonumber
\end{align}
Problem \eqref{slot_prob_local_comp} is convex, and its closed form solution, which can be easily derived via the Karush-Kuhn-Tucker conditions \cite{boyd2004convex}, reads as follows:
\begin{equation}
    f^{l*}_t= \min\bigg(\max\bigg(\sqrt[\leftroot{-1}\uproot{4}\scriptstyle 4]{\frac{(Y_t+Z_t) w_t}{3V\kappa}}, f_{\min}^l\bigg), f^l_{\rm max}\bigg)
\end{equation}
\subsection{Communication sub-problem}
The problem to be solved for the uplink power allocation over subcarriers and the RIS configuration is the following
\begin{align}\label{slot_prob_comm}
      \hspace{0 cm}\min_{\theta}& \;V\sum\nolimits_{b\in\mathcal{B}}p_{b,t} + \left(Y_t+Z_t\right)A_t/R_t\\
   &\textrm{subject to}\quad (a)\text{-}(f) \;\text{of}\;\eqref{problem_form}\nonumber
\end{align}
with $\theta=[\{p_{b,t}\}_b,\{f_{n,t}\}_{n},\{S_{n,t}\}_{n}, \{\chi_{n,t}\}_{n}]$, with $R_t$ defined as in \eqref{data_rate}. Problem \eqref{slot_prob_comm} is still a mixed-integer non-convex program. However, given an RIS configuration, the problem is convex and can be efficiently solved \textit{optimally} through efficient procedures such as interior-point methods \cite{boyd2004convex}. Therefore, we now propose an efficient heuristic to optimize the RIS configuration. 
The algorithm to optimize the RIS builds on the one proposed in \cite{di2022dynamic}, in which RIS elements responses are subsequently selected with the goal of increasing a weighted sum of channel power gains. However, \cite{di2022dynamic} is limited to frequency selective channels and focuses on a multi-user case. Therefore, in this case, we propose to subsequently select RIS parameters (i.e. $S_{n,t}$, $f_{n,t}$, and $\chi_{n,t}$) for each element, in order to maximize the sum of the channel power gains over all subcarriers. In particular, as reported in Algorithm \ref{alg:RIS}, the greedy procedure starts from all elements with $S_n=0$, i.e. an RIS with no reflections. Starting from this condition, the sum of all channel gains is computed (S.$1$). Then, going through all elements, each elements' response, encoded by $f_{n,t}$, $\chi_{n,t}$, and $S_{n,t}$, is chosen to maximize the sum of channel gains $\Delta=\sum_{b\in\mathcal{B}}\alpha_t(f_b)$, given the other elements' parameters $f_{-n,t}$, $\chi_{-n,t}$, and $S_{-n,t}$ (S.$2$), choosing $S_{n,t}\in[0,1]$ to guarantee constraint $(d)$ of \eqref{problem_form}. The described procedure obviously guarantees a monotone non-decreasing behaviour of $\Delta$, i.e. the sum channel power gain.
\begin{algorithm}[t]
\SetAlgoLined
\textbf{Input:} Channels' frequency responses \smallskip

(S.$1$) Set $S_n=0$ $\forall n$, $\Delta=\sum_{b\in\mathcal{B}}\alpha_t(f_b)$.\smallskip\\
\For{$n=1:N$}{
(S.$2$) $\{\bar{f}_n,\bar{\chi}_n\}_{n}=\!\!\!\!\!\!\!\underset{\{f_n\in\Omega,\chi_n\in\mathcal{X}\}}{\textrm{arg}\max}\Delta(\bar{f}_n,\bar{\chi}_n;\bar{f}_{-n},\bar{\chi}_{-n})$,\\ with $\bar{S}_n=(\underset{f_b}{\textrm{arg}\max}\;|\phi(f_b;\bar{f}_n,\bar{\chi}_n)|)^{-1}$
}
Set $\{f_n=\bar{f}_n\}_n$, $\{\chi_n=\bar{\chi}_n\}_n$, $\{S_n=\bar{S}_n\}_n$\\
\textbf{Output:} $\{f_n\}_n,\{\chi_n\}_n,\{S_n\}_n$
\caption{\textbf{Greedy RIS optimization}} 
\label{alg:RIS}
\end{algorithm}
\section{Numerical results}
\begin{figure}
    \centering
    \includegraphics[width=.95\columnwidth]{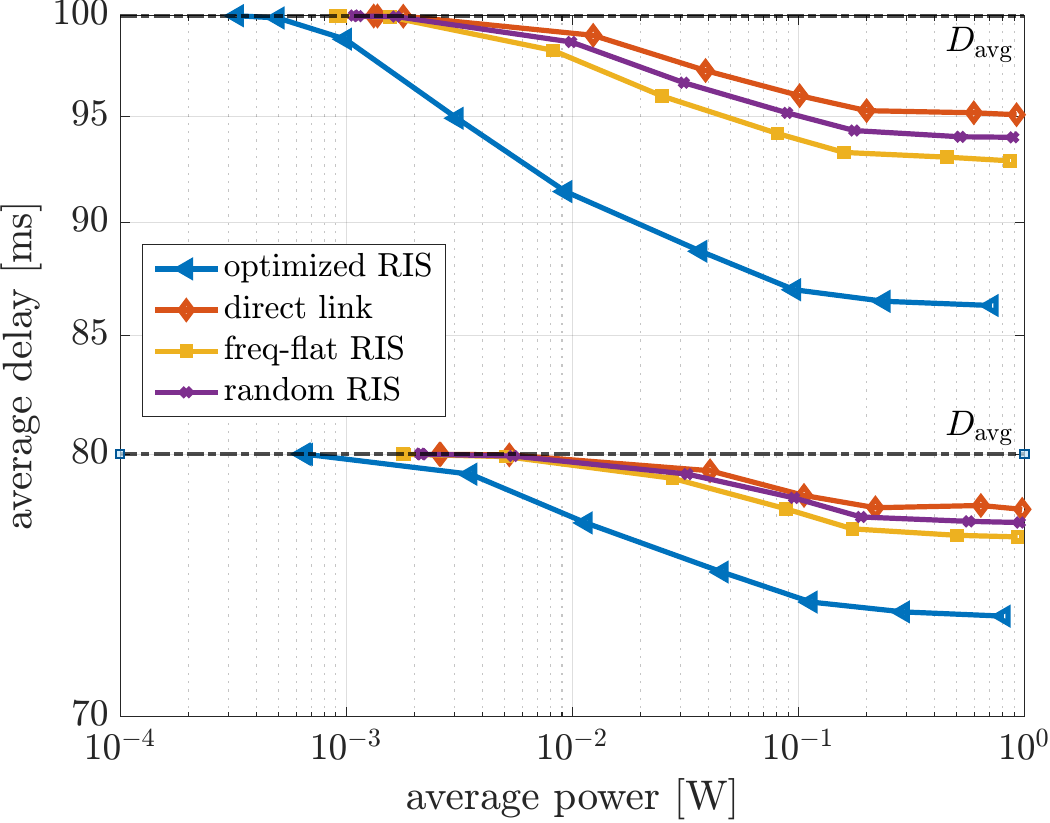}
    \caption{Average delay vs. average power consumption.}
    \label{fig:tradeoff}
\end{figure}
\begin{figure}
    \centering
    \includegraphics[width=.95\columnwidth]{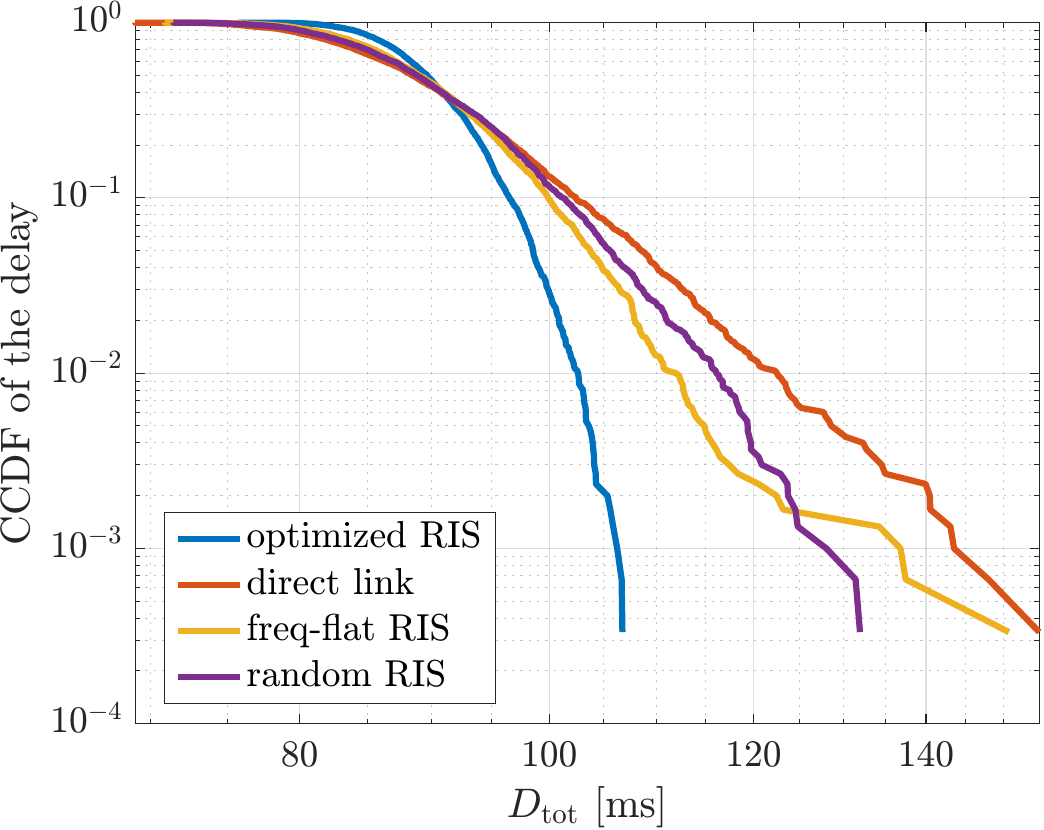}
    \caption{Survivor function of the E2E delay.}
    \label{fig:CCDF1}
\end{figure}
\begin{figure}
    \centering
    \includegraphics[width=.95\columnwidth]{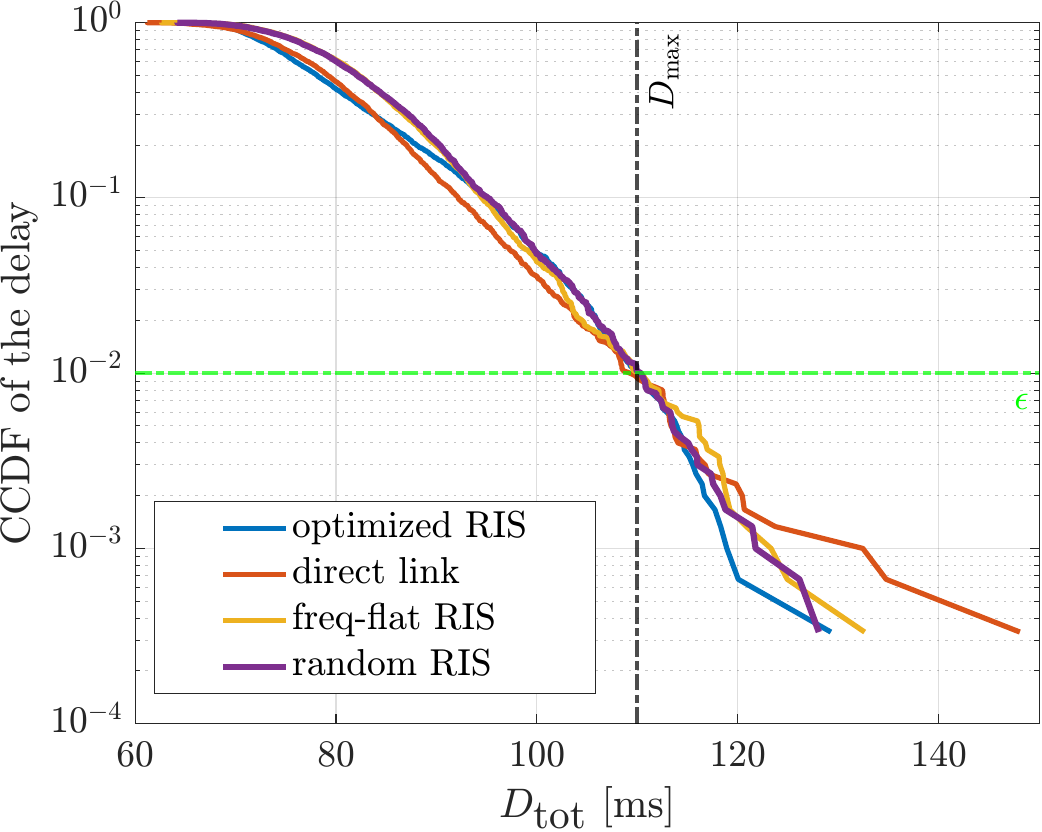}
    \caption{Survivor function with probabilistic constraints.}
    \label{fig:CCDF2}
\end{figure}
\begin{figure}
    \centering
    \includegraphics[width=\columnwidth]{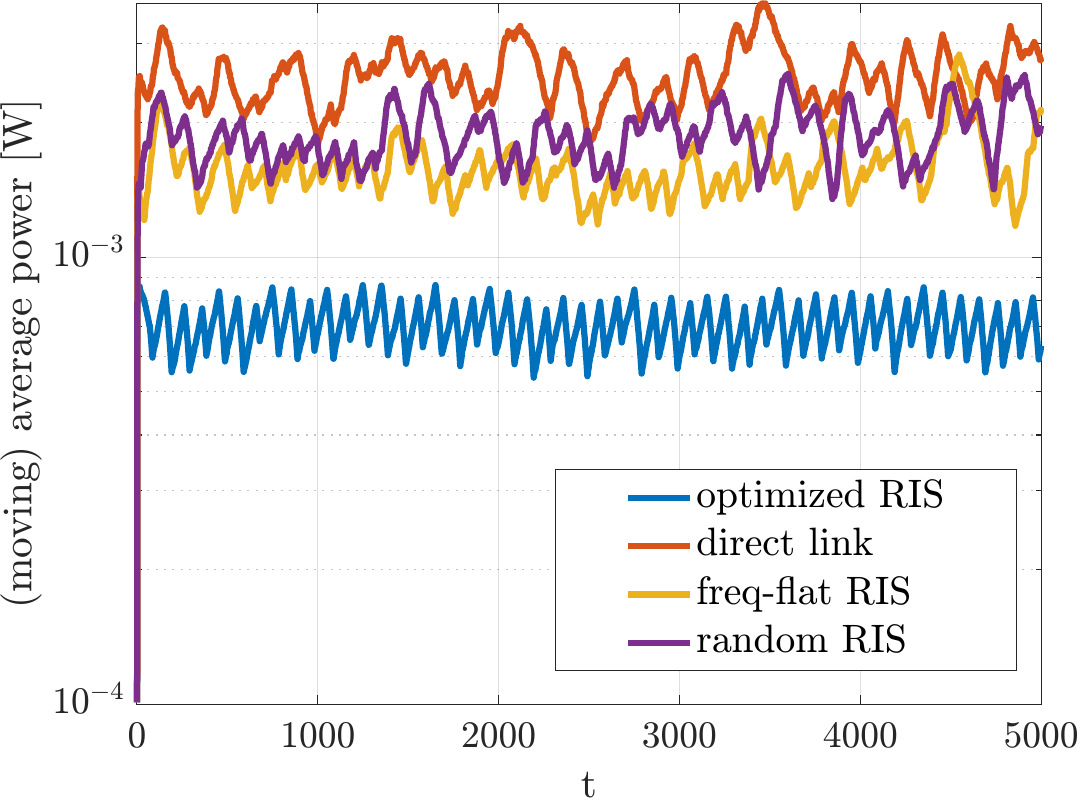}
    \caption{Average power consumption vs. time-slot index.}
    \label{fig:power_vs_t}
    \vspace{-.2 cm}
\end{figure}

\begin{table}[!t]
\begin{center}
\caption{Simulation Parameters}
\label{tab1}
\begin{tabular}{| c | c |}
\hline
\hline
\textbf{Parameters} & \textbf{Values}\\
\hline 
\hline
Device's transmit power ${p_{\min}}$, ${p_{\max}}$ (mW)& $0.1,\,100$\\
\hline
Carrier frequency ${f_c} $ (GHz) &$3.5$\\ 
\hline
Subcarrier spacing ${W} $ (MHz)& $1$\\
\hline
number of subcarriers ${B}$ & $16$\\
\hline
Noise power spectral density ${N_0} $ (dBm/Hz)& $-174$\\
\hline
Server's CPU maximum frequency ${f_{\max}} $(GHz) & $10$\\
\hline
Device's CPU frequencies ${f^l_{\min}}$, ${f^l_{\max}}$ (GHz)& $0.01, \, 1$\\
\hline
Processor's effective switched capacitance ${\gamma}$ & $10^{-27}$\\
\hline 
\hline
\end{tabular}
\end{center}
\vspace{-.7 cm}
\end{table}

In this section, we assess the performance of our proposed optimization method. The parameters used for the simulations are reported in Table \ref{tab1}. In addition to those, the amount of local computational demand $w^l_t$, the amount of data $A_t$ transmitted by the user, and the remote computational demand $w^r_t$ are generated from Poisson distributions with mean $\overline{w}^l = 5 \times 10 ^5$ CPU cycles, $\overline{A} = 2$ Mb and $\overline{w}^r = 5\times 10^7$ CPU cyles, respectively. We consider a 2D scenario composed by one device placed at $(10,30)$, one RIS of $N=100$ elements placed at $(-5,2.5)$ and an AP placed at $(0,0)$. At each time slot $t$, Rayleigh fading frequency selective SISO channels are generated, with $L=4$ delayed taps, and their Fourier transform is performed to obtain the channel transfer functions $\mathbf{h}_t$, $\mathbf{g}_t$ and  $h^{\textrm{LoS}}_t$ (cf. \eqref{CNR}). Typical frequency dependent pathloss is considered, with exponent $2$, $2$ and $4$ for the distance, respectively. Regarding constraints $(b)$ and $(c)$ of \eqref{problem_form}, we choose $\Omega$ equal to the set of subcarriers $\mathcal{B}$, and $\chi=\{10,25,50,100\}$. In all simulations, we compare the RIS-aided performance with the Lorentzian model and our frequency selective-aware optimization, termed as \textit{optimized RIS}, with three benchmarks: i) the case without the RIS, termed as \textit{direct link}, with resources optimized with the proposed method; ii) the frequency flat RIS case, termed as \textit{freq-flat RIS}, in which the RIS response does not follow the realistic Lorentzian model, but it is flat across all frequencies. In this last case, the RIS is optimized as in \cite{di2022dynamic}, while resources are optimized through our method; iii) \textit{random RIS}: the RIS parameters are randomly selected from the feasible set at each time slot, and resources are accordingly optimized with our method. 
As a first result, in Fig. \ref{fig:tradeoff} we show the trade-off between the average E2E delay and the power consumption, obtained by increasing the parameter $V$ (cf. \eqref{DPP}) from right to left. Two different average delay constraints are considered, namely $100$ ms and $80$ ms, while no probabilistic delay constraint is imposed in this simulation. As we can notice, for all curves, the average E2E delay increases as the power consumption decreases, until approaching the delay constraint. However, the \textit{direct link} case exhibits the worst performance in terms of delay-power trade-off, as it achieves higher delays for a fixed power consumption. At the same time, the best performance is achieved by the \textit{optimized RIS} (i.e. the proposed method).
To make a fair comparison, let us focus on the highest values of $V$, and therefore the highest delay (points on the left side of the plot), which is indeed the service requirement. Here, we can notice how the \textit{optimized RIS} case achieves the lowest power consumption, while a slight gain with respect to the \textit{direct link} case is achieved by the \textit{freq-flat RIS} and the \textit{random RIS}. However, the last two benchmarks do not achieve the considerable gain achieved by the proposed method, due to the awareness of the latter on frequency selective channels and RIS responses.\\ 
\indent Let us notice that Fig. \ref{fig:tradeoff} only shows the average E2E delay, while it ignores the distribution of the delay. To this end, in Fig. \ref{fig:CCDF1}, we plot the complementary cumulative distribution function (CCDF) of the delay, for the highest value of $V$, i.e. the left points of Fig. \ref{fig:tradeoff}, for $D^{\textrm{avg}}=100$ ms, for all cases. From this figure, we notice an even more interesting outcome: the \textit{optimized RIS} exhibits much better performance in terms of delay distribution, while the three benchmarks present longer tail distributions, i.e. higher delay variability, also due to the fact that the RIS improves link reliability and no probabilistic constraints have been imposed in this simulation. Then, we run a simulation with $D^{\textrm{avg}}=100$ ms, $D^{\max}=110$ ms, and $\epsilon=10^{-2}$. We show the resulting survivor function in Fig. \ref{fig:CCDF2}, validating the ability of our algorithm to guarantee the probabilistic constraint, as shown by the intersection of the curves with $D^{\max}$ and $\epsilon$. However, given these comparable distributions, it is fundamental to plot the power consumption, shown in Fig. \ref{fig:power_vs_t} for all cases. Again, despite the fact that all cases exhibit the same performance in terms of delay distribution, the power consumption of the proposed \textit{optimized RIS} method is considerably lower than the others, while \textit{freq-flat RIS} and \textit{random RIS} show again a lower gains as compared to the \textit{direct link} benchmark case, which exhibits indeed the worst performance due to its sensitivity to channel fading conditions.
\vspace{-.2 cm}
\section{Conclusions}
We proposed a first step towards the optimization of radio and computing resource over frequency selective
RIS-aided communication channels, with a recently developed frequency dispersive RIS response model. We formulated and solved a problem taking into account short-term constraints involving local computing, transmit power and RIS parameters, as well as long-term constraints involving average and probabilistic E2E delays. Through numerical simulations, we assess the performance of our method, especially with respect to benchmark solutions that do not involve RISs or do not perform a frequency selective-aware optimization. We believe the results shown in this work pave the way to future investigations on frequency selective RIS-aided MEC in more complex scenarios, to finally assess the benefits of such technology in future connect-compute 6G and beyond networks.
\vspace{-.1 cm}

\bibliographystyle{IEEEtran}
\bibliography{IEEEabrv, references}

\end{document}